\documentclass[preprint,12pt]{elsarticle}
\usepackage{amsmath}
\usepackage{graphicx}

\begin{document}

\begin{frontmatter}
\title{First-principles study of spin-wave dispersion in Sm(Fe$_{1-x}$Co$_{x}$)$_{12}$}

\author[1,3]{
Taro Fukazawa
}
\author[2,3]{
Hisazumi Akai
}
\author[1,3]{
Yosuke Harashima
}
\author[1,3]{
Takashi Miyake}
\address[1]{
National Institute
of Advanced Industrial Science and Technology,
Tsukuba, Ibaraki 305-8568, Japan}
\address[2]{
The Institute for Solid State Physics,
The University of Tokyo,
Kashiwano-ha, Kashiwa,
Chiba 277-8581, Japan}
\address[3]{
ESICMM, National Institute for Materials Science,
Tsukuba, Ibaraki 305-0047, Japan}

\begin{abstract}
We present spin-wave dispersion in Sm(Fe$_{1-x}$Co$_x$)$_{12}$ calculated
based on first-principles.
Anisotropy in the lowest branch of the spin-wave dispersion
around the $\Gamma$ point is discussed.
Spin-waves propagate more easily along $a^*$-axis than along
$c^*$-axis, especially in SmFe$_{12}$.
We also compare values of the
spin-wave stiffness with those obtained from an experiment.
The calculated values are in good agreement
with the experimental values.
\end{abstract}


\end{frontmatter}

\section{Introduction}
Magnetic properties at finite temperatures are important in applications of hard magnets, and ab initio modeling for spins has become one of the standard techniques today.
The spin-wave dispersion can be derived from such a model. It offers intuitive description of magnetic collective modes, which is important in understanding finite temperature properties. It is also possible to compare the dispersion directly with that obtained by experiments.

Magnetic compounds with the ThMn$_{12}$ structure have regained attention since their potential as the main phase in a hard magnet was reevaluated by a first-principles study \cite{Miyake14} and experimental works \cite{Hirayama15, Hirayama15b} in these years. Hirayama et al. have recently synthesized Sm(Fe$_{1-x}$Co$_{x}$)$_{12}$ films for $x = 0, 0.1, 0.2$, and shown that Sm(Fe$_{0.8}$Co$_{0.2}$)$_{12}$ has favorable magnetic properties, including the spontaneous magnetization of 1.78 T at room temperature \cite{Hirayama17}. 

In this paper, we  present spin-wave dispersion in Sm(Fe$_{1-x}$Co$_{x}$)$_{12}$ calculated
based on first-principles for $x=0$ and $0.2$.
Because there is no experiment clarifying its
spin-wave dispersion to the best of our knowledge, we compare values of the
spin-wave stiffness with those obtained from the experiment by Hirayama
et al.\cite{Hirayama17}
We also discuss anisotropy in the lowest branch around the $\Gamma$
point: spin-waves propagate more easily along $a$-axis than along
$c$-axis especially in SmFe$_{12}$.

\section{Methods}
\begin{figure}[!t]
\centering
 \includegraphics[width=4in,bb=0 0 1200 1050]{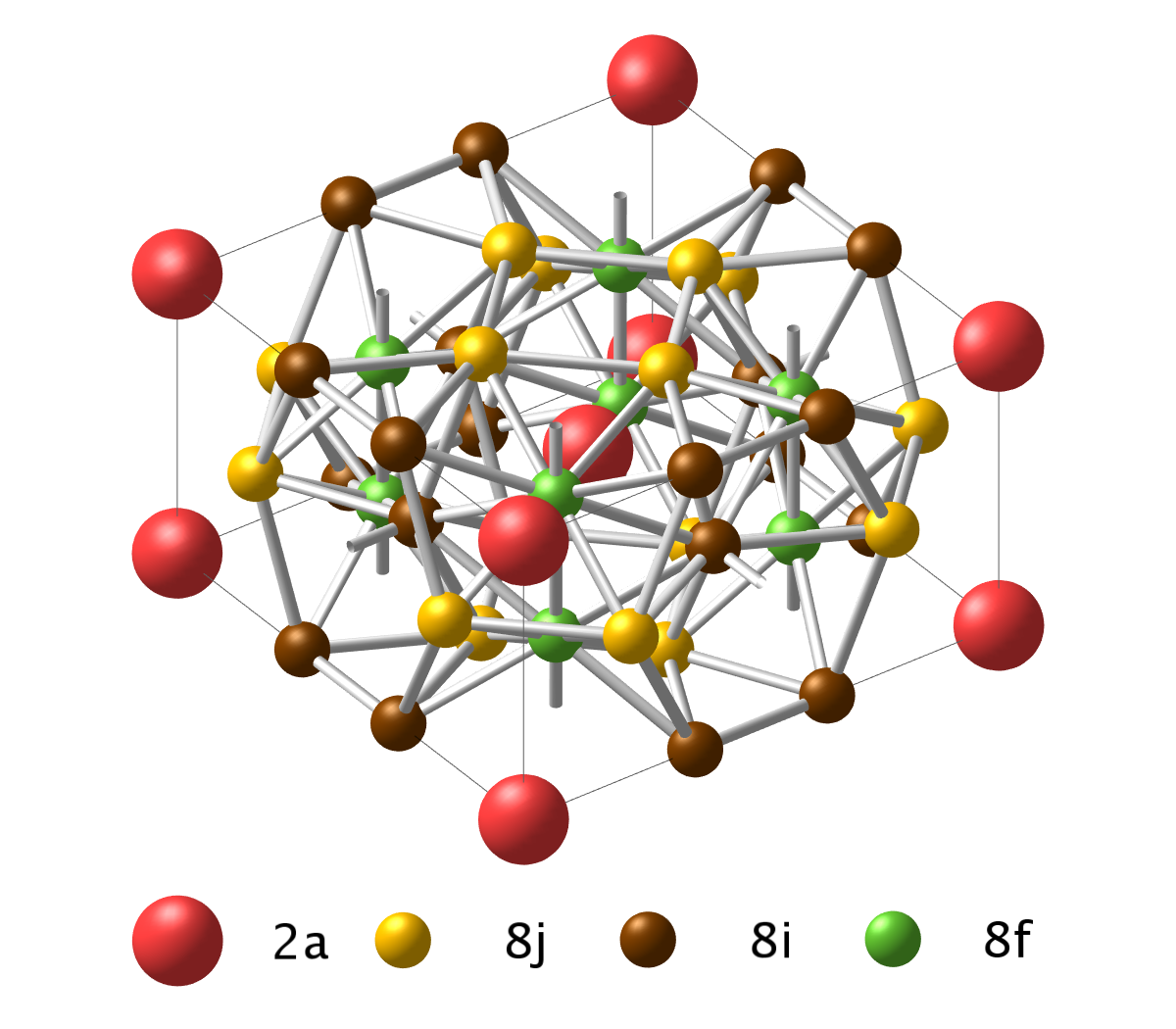}
\caption{Crystal structure of the ThMn$_{12}$ structure. The atomic positions indicated by the Wyckoff positions (2a, 8j, 8i, 8f) are shown. Some bonds are shown to serve as eye-guides.
}
\label{crystalstructure}
\end{figure}

We use the Korringa-Kohn-Rostoker Green function method for solving
the Kohn-Sham equation of density functional theory \cite{Hohenberg64,Kohn65}.
The
exchange-correlation functional is approximated within the local density
approximation \cite{Kohn65}.
The f-orbitals at the Sm site are treated as a trivalent
open core with the spin-configuration limited by Hund's rule,
and the self-interaction correction is applied to the
orbitals.
The spin-orbit coupling is disregarded in the calculation
except that the effect is implicitly taken into account in
the spin-configuration of the f-electrons.
We assume the Fe and Co atoms randomly occupy the 8f, 8i and
8j site in the ThMn$_{12}$ structure
[Space group: $I4/mmm$ (\#139); see also Figure \ref{crystalstructure}]
and their site preference is
disregarded.
In the calculation of spin-wave dispersion, 
we treat the randomness with the virtual crystal
approximation (VCA).
However, for the 
the other part of the calculation, namely,
the magnetization and the Curie temperature,
we use the coherent potential approximation (CPA),
which is more sophisticated than VCA concerning the randomness.
To compare the CPA results with that of VCA, we calculate 
the magnetization and the Curie temperature also within VCA.
The reason why we do not use CPA in the calculation of 
the spin-wave dispersion is addressed later in this section.
The experimental lattice
constant $a$ and $c$ given in Ref. \cite{Hirayama17}
are used in those calculations.
We use the calculated values of
the inner parameters for SmFe$_{12}$ given in Ref. \cite{Fukazawa17}.

The magnetic coupling is calculated using
Liechtenstein's formula \cite{Liechtenstein87}.
In formalism, we use those values as $J^{\mu,\nu}_{i,j}$ in the following classical 
Heisenberg Hamiltonian:
\begin{equation}
 H = - \sum_{i,\mu} \sum_{j,\nu}
  J^{\mu,\nu}_{i,j} \,
  \vec{e}^{\,\mu}_i \cdot \vec{e}^{\,\nu}_j
\label{Heisenberg_1}
\end{equation}
where $\vec{e}^{\,\mu}_i$ is a unit vector 
that is in the direction of the magnetic moment 
at the $\mu$th site 
in the $i$th unit cell.

In calculation of spin-waves, we consider small fluctuation of 
$\vec{e}$ from the alignment in the ground state,
$\vec{e}^{\,\text{GS}, \mu}_i$.
We assume that $\vec{e}^{\,\text{GS}, \mu}_i$ is parallel or antiparallel to 
the $z$-direction.
Elementary excitations of spin-waves
$\vec{e}^{\,\mu}_i=\vec{e}^{\,\text{GS}, \mu}_i + \vec{u}^{\,\mu}(\vec{q})\exp{(i\omega t - i\vec{q}\cdot\vec{R}_i )}$
can be obtained by diagonalizing a matrix $\mathcal{J}(\vec{q})$
constructed from 
$J^{\mu,\nu}_{i,j}$,
where $u^{\,\mu}_{x}(\vec{q}),u^{\,\mu}_{y}(\vec{q})\ll 1$ and $u^{\,\mu}_{z}(\vec{q})\simeq 0$
are assumed
for the $x$-, $y$- and $z$-component, respectively.
We summarized derivation of $\mathcal{J}(\vec{q})$ in \ref{Appendix_1}
without considering an equation of motion.
We refer readers to Ref. \cite{Halilov98} for its relation to the dynamics.

In Liechtenstein's formula, spin-rotational perturbations
at the $i$ and $j$ site are considered, and
the excitation energy is interpret as 
intersite magnetic interaction $J^{\mu,\nu}_{i,j}$
\cite{Liechtenstein87}.
The formula consists of the perturbation of the local potentials and 
the scattering path operator. 
The scattering path operator can be obtained from the Green's function of
the Kohn-Sham system, which is usually obtained as a function in 
the reciprocal space.
While the spin-rotational perturbation can be formally transformed into the reciprocal space (because it is Kronecker-delta-like), 
it is difficult to transform
the scattering path operator into the real space 
without loss of precision.
Therefore, it is advantageous to construct 
$\mathcal{J}(\vec{q})$ in the reciprocal space.
Pajda et al has also pointed out that this type of direct calculation is 
possible \cite{Pajda01}.

Based on the Fourier transformed Liechtenstein's formula,
we have developed a method for
obtaining the $\omega$--$\vec{q}$ dispersion
directly in the reciprocal space that is combined with the KKR method
for this study.
In the present calculation,
$16 \times 16 \times 16$ q-points in the Brillouin zone are considered.
As a post-process,
we interpolate quantities for arbitrary $\vec{q}$ vectors using the values on the $q$-mesh
as follows.
We first construct a Fourier series that reproduces all the values on the mesh.
This includes, however, terms with high frequencies that 
cannot be accurately determined with the $q$-mesh.
We truncate all these terms (low-pass filtering),
and add a constant so that 
the quantity at the $\Gamma$ point becomes identical
to the original one.

The reason why we use VCA instead of CPA in the calculation of spin-wave dispersion is the following.
Liechtenstein's formula (with CPA) gives $J$ that also depends on the two elements at the ends.
We can use those $J$'s to construct the Heisenberg model with a random configuration of elements with the justification described in \cite{Buczek16}.
Then, we have to consider the sample average for the spin-wave dispersion.
It needs too large a supercell to take the average directly when $J$ is long-ranged. Although application of CPA to the Heisenberg model was proposed in \cite{Buczek16} to overcome this difficulty,
the expected resource consumption is still too high for the systems we considered.
It is because the calculation
needs a fine mesh in the Brillouin zone
in order
to obtain spin-wave dispersion
around the $\Gamma$ point,
which is of particular interest here.
On this ground, we abandoned using $J$ that depends on the atomic species.
In order to obtain such averaged $J$, VCA would be enough.

\section{Results and discussion}
\begin{figure}[!t]
\centering
 \includegraphics[width=3.5in,bb=0 0 504 360]{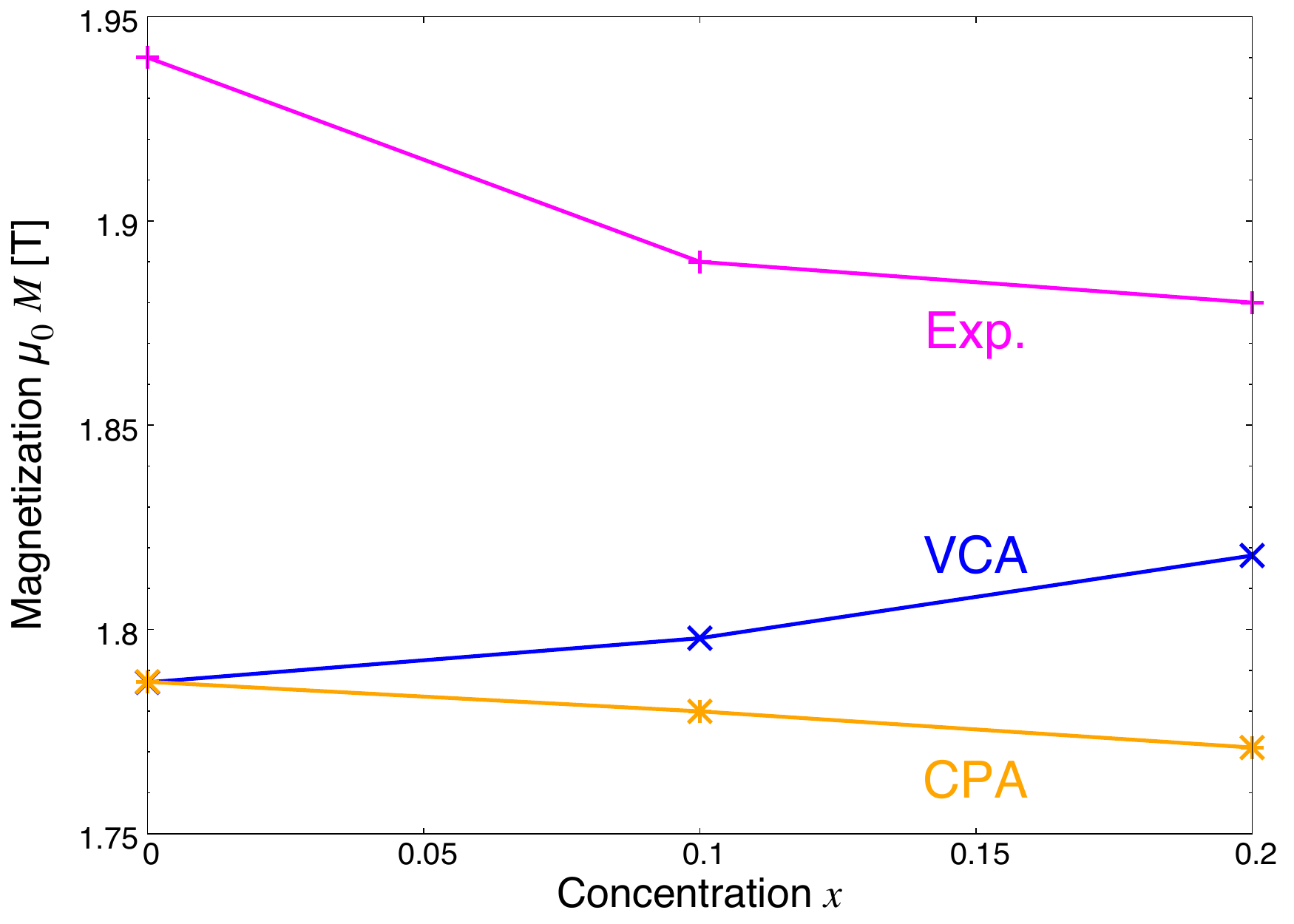}
\caption{Values of magnetization calculated with VCA and CPA, and those from an experiment (extrapolated to 0 K) \cite{Hirayama17} as functions of concentration $x$.}
\label{fig_vs_cpa}
\end{figure}

We first present our results for magnetization within VCA and CPA,
and compare them with the experimental values in Ref. \cite{Hirayama17}.
Figure \ref{fig_vs_cpa} shows the values of the magnetization as functions of $x$ (the Co concentration).
The contribution from f-electrons at Sm sites are included
in those values assuming that the f-electrons have magnetic moment
of $g_J\sqrt{J(J+1)}=0.85 \mu_\mathrm{B}$, which adds approximately 0.05 T to the magnetization.

The theoretical prediction reasonably agrees with the experimental values
with the underestimation of 0.15 T at worst.
In a previous first-principles study using a PAW-GGA method
(which consider the spin-orbit coupling only in the spin-configuration
at the Sm sites as in our calculation),
magnetization of SmFe$_{12}$ was evaluated as 1.83 T \cite{Harashima14b},
which is not much different from the present value.
The deviation from the experiment is partly attributed to orbital moments.
When we considered the spin-orbit coupling at the Fe sites, 
the gain was 0.03 T in SmFe$_{12}$.

The experimental observation that
the Co introduction reduces
the magnetization
at low temperature
is theoretically understandable.
It has already been discussed that the optimal percentage
of substitution of Co can significantly depend
on how the host system has a room for improvement
in the magnetic moment \cite{Harashima16}.
In the case of SmFe$_{12}$, the Co substitution
can slightly enhance the ferromagnetism due to hybridization
between Fe and Co, however, this does not overcome
the expansion of the volume.
This reduction is well reproduced by the CPA calculation.

As for validity of VCA,
it is adequate in the sense that 
the deviations of the values from CPA
are within a few percent.
We should note, however, that
the tendency as a function of $x$ is opposite.
In VCA, the enhancement of the ferromagnetism with increasing $x$
is exaggerated
by totally forgetting the inhomogeneity caused by the randomness.

\begin{figure}[!t]
\centering
 \includegraphics[width=3.5in,bb=0 0 504 360]{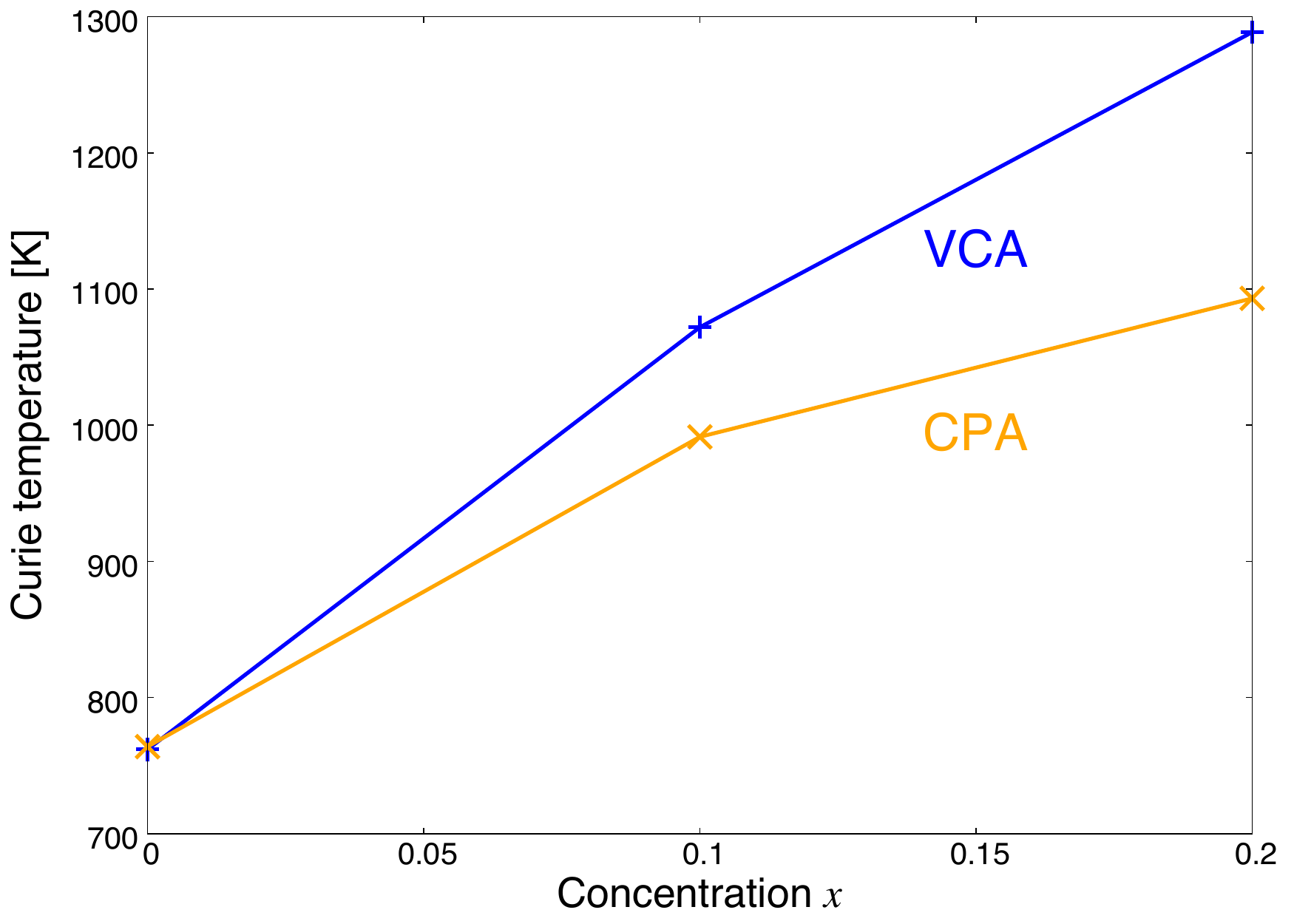}
\caption{Calculated values of Curie temperature within VCA
and CPA
as functions of concentration $x$.
In both cases,
the mean-field approximation is used.
}
\label{fig_vs_cpa2}
\end{figure}

\begin{figure}[!t]
\centering
 \includegraphics[width=3.5in,bb=0 0 504 360]{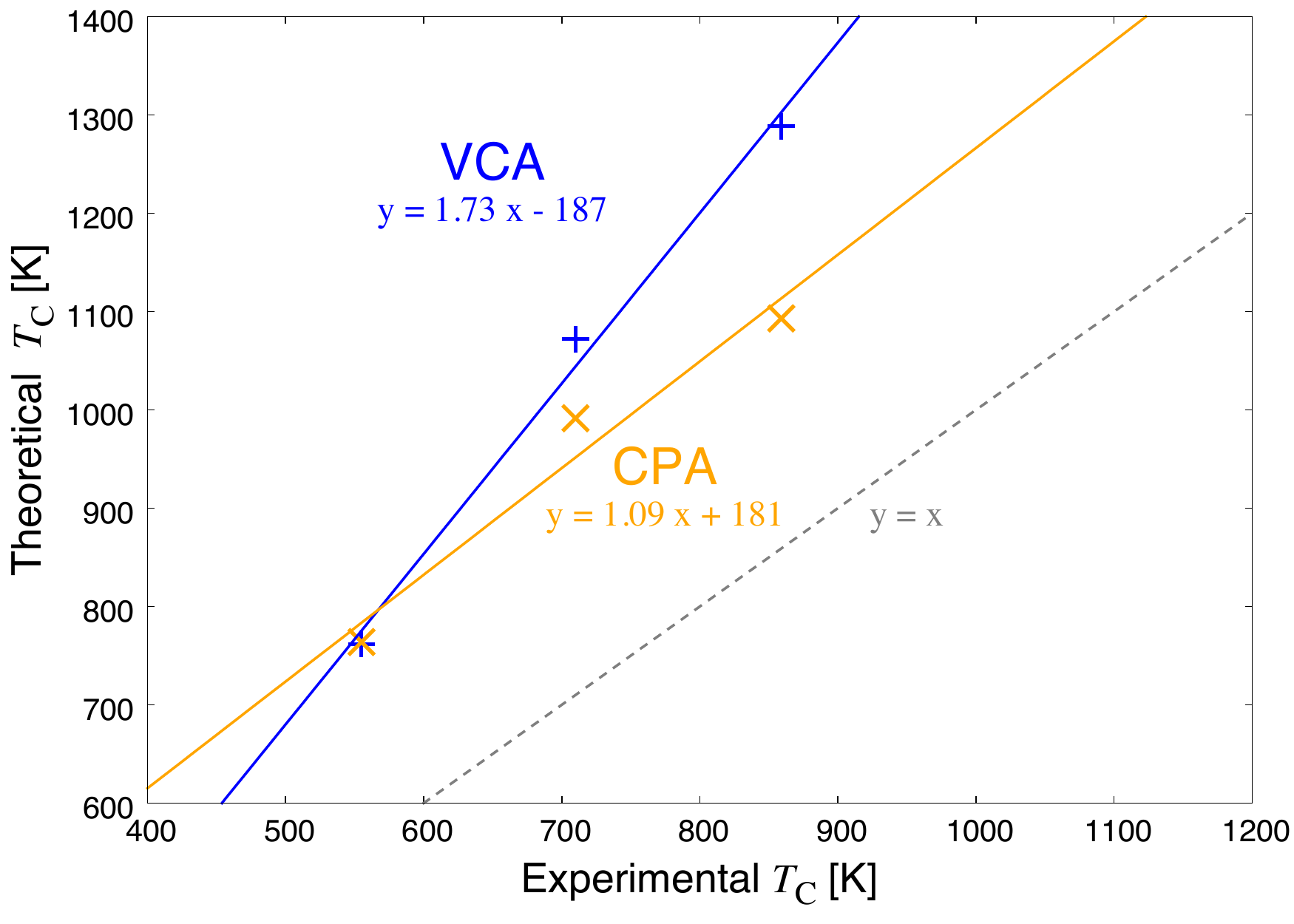}
\caption{Experimental values (horizontal) of Curie temperature \cite{Hirayama17} versus 
theoretical values (vertical) within VCA and CPA.
In both cases,
the mean-field approximation is used.
Results of linear regression are also shown in the figure}
\label{fig_vs_exp}
\end{figure}

Figure \ref{fig_vs_cpa2} shows values of the calculated
Curie temperatures as functions of $x$ within VCA and CPA.
In both cases, we use the mean-field approximation (MFA)
as described in Ref. \cite{Fukazawa18}.
Those Curie temperatures ($T_\mathrm{C}$) are overestimated
due to the use of MFA, however,
it reproduces relative change of $T_\mathrm{C}$ with respect to $x$
in experiments
(see also our previous study \cite{Fukazawa18}).
Figure \ref{fig_vs_exp} compares the numerical results with the
experimental Curie temperature \cite{Hirayama17}.
Results of linear regression are also shown in the figure.
The relative change of $T_\mathrm{C}$ is reproduced
well in CPA, which can be seen from 
the value of the gradient (1.09) being close to 1.
In VCA, 
the tendency toward ferromagnetism is again excessive, however,
the values of the Curie temperature is significantly correlated 
with the experimental values. We therefore expect that VCA
also offers informative description of Sm(Fe$_{1-x}$Co$_x$)$_{12}$.

\begin{figure}[!t]
\centering
\includegraphics[width=3.5in,bb=0 0 504 360]{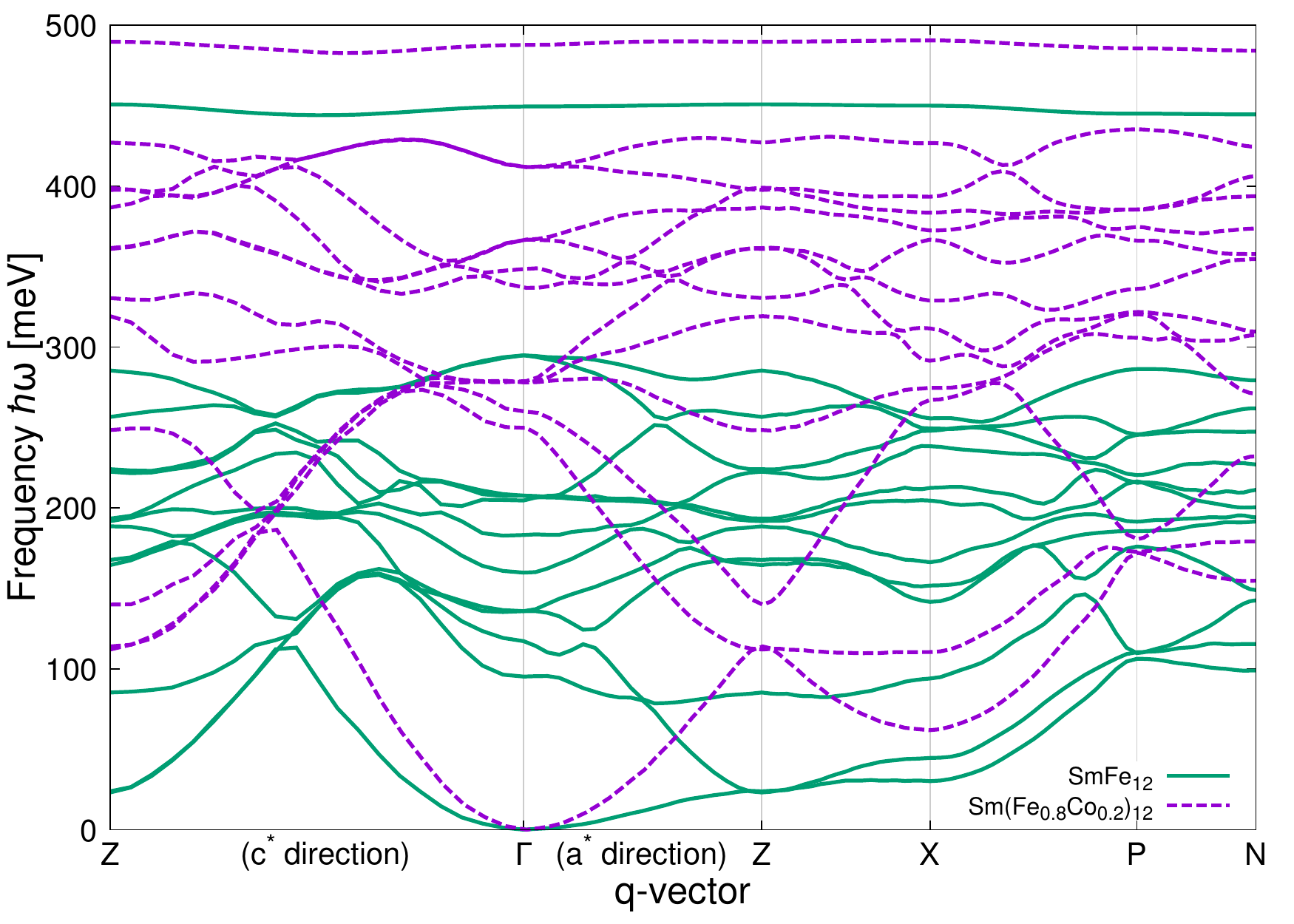}
\caption{The spin-wave dispersion ($\hbar\omega$ as a function of $\vec{q}$)
in SmFe$_{12}$ (green) and Sm(Fe$_{0.8}$Co$_{0.2}$)$_{12}$ (purple).}
\label{fig_sim}
\end{figure}

We then present our results for the spin-wave dispersion.
Figure \ref{fig_sim} shows that in SmFe$_{12}$ (solid lines) and Sm(Fe$_{0.8}$Co$_{0.2}$)$_{12}$ (broken lines). The broken lines are aligned upward compared to the solid lines.
The introduction of Co in Sm(Fe$_{0.8}$Co$_{0.2}$)$_{12}$ enhances the magnetic exchange interaction and makes the excitation energy for the collective modes higher than in SmFe$_{12}$. 

In SmFe$_{12}$, the
curvature of the lowest branch around the $\Gamma$ point along $c^*$-axis ($D_c$) is larger 
than that along $a^*$-axis ($D_a$).
The smallness of $D_a$ indicates that 
spin-waves propagates more easily along 
$x$-axis than $z$-axis.
The value of $\omega$ 
at Z [= (1\,0\,0)]
is conspicuously small,
which characterizes the smallness of $D_a$.
This anisotropy originates from the tetragonal asymmetry
of the system, which has two effects:
(1) it elongates the Wigner-Seitz cell in the reciprocal space
along $c^*$-axis in the case of $c < a$,
and (2) allows the distribution of 
$J^{\mu,\nu}_{i,j}$ to be anisotropic.
However, the elongation of the Wigner-Seitz cell
makes $D_a/D_c$ larger
when the values of $J^{\mu,\nu}_{i,j}$ are kept unchanged
(see also \ref{Appendix_C}).
Therefore, the smallness of $D_a$ must be attributed to 
the anisotropic distribution of $J^{\mu,\nu}_{i,j}$,
which would be closely related to the structure of the network of
the transition atoms. 
As for Sm(Fe$_{0.8}$Co$_{0.2}$)$_{12}$, the
curvatures become more isotropic due to the enhancement of the exchange
interaction.

The calculated value of the curvatures 
from the dispersion
are shown in Table \ref{table}.
Those are determined by fitting of a quadratic function
to the data points within a $0.2\times \frac{2\pi}{a}$ radius
of the $\Gamma$ point.
The spin-wave stiffness $D$ corresponding to the coefficient of $T^{3/2}$
in the Bloch's theory of the isotropic case
is also determined in the following way.
Consider transformation
$(q_a, q_b, q_c)\rightarrow (\sqrt{D/D_a} q'_a, \sqrt{D/D_a}q'_b, \sqrt{D/D_c} q'_c)$,
where $q$'s are the components of $\vec{q} \equiv (q_a, q_b, q_c)^{\mathrm T}$.
This transforms the approximate harmonic dispersion to an isotropic one:
$\hbar\omega \approx D_a q_a^2 + D_a q_b^2 + D_c q_c^2 \rightarrow D ({q'}^{2}_a + {q'}^{2}_b + {q'}^{2}_c )$.
One should then pay attention to the change of the state density in the $q'$ space.
When the Jacobian $\sqrt{D^3/(D_a^2 D_c)}$ is one the state density is kept unchanged.
Therefore, $D = (D_{a}^2 D_{c})^{1/3}$ gives the corresponding spin-stiffness.

The calculated values in Table \ref{table} are in
good agreement with the value obtained by Hirayama et al \cite{Hirayama17}
for all the $x$'s.
The theoretical $D$ tends to be smaller than the experimental 
values, which is conspicuous especially at $x=0$.
For a finite $x$,
this underestimation is partly and accidentally canceled by
the excessive tendency toward ferromagnetism in VCA.

\begin{table}[!t]
\renewcommand{\arraystretch}{1.3}
\caption{Calculated values of the spin-wave stiffness for
 Sm(Fe$_{1-x}$Co$_{x}$)$_{12}$ compared with the experimental values by Hirayama et al
 \cite{Hirayama17}. The curvatures around the $\Gamma$ point in the lowest branch and
 along the $a^*$- and $c^*$-axis are also shown as $D_a$ and $D_c$. The stiffness
 values are in the unit of $\mathrm{meV \AA}^2$.}
 \label{table}
 \centering
 \begin{tabular}{rcccc}
  \hline
  \hline
  $x$  & $D$ ($D_{\rm Exp.}$) & $D_{a}$ & $D_{c}$ \\
  \hline
    0 & 118  (179) &  91.6 & 194  \\
  0.1 & 252  (251) & 220   & 332  \\
  0.2 & 324  (351) & 290   & 407  \\
 \hline
\hline
 \end{tabular}
\end{table}

The values of $D_a$ and $D_c$ can be converted to
the exchange stiffness, $A$.
We here consider the following macroscopic Hamiltonian
for magnetization $\vec{M}$ and the magnetic coupling:
\begin{equation}
 E = 
  -\int_{\Omega} dx\,dy\,dz \,
  \vec{m}^\mathrm{T}
  \left(
     \vec{\nabla}^\mathrm{T} \mathcal{A} \vec{\nabla}
  \right)
  \vec{m},
\end{equation}
where
$\vec{m}\equiv \vec{M}/|\vec{M}|$,
$\vec{\nabla}=(\partial_x, \partial_y, \partial_z)^{\mathrm T}$,
$\Omega$ denote the domain,
and $\mathcal{A}$ is a $3 \times 3$ constant tensor that is 
real, symmetric and positive definite.
We also assume that
the absolute value of the magnetization, $|\vec{M}|$, is constant through the space.
Our Hamiltonian is identical to that used by
Belashchenko \cite{Belashchenko04}
except for the surface term with a partial integration.

Let us consider
$\mathcal{A}=\text{diag}[A_x, A_x, A_z]$
for 
the Sm(Fe$_{1-x}$Co$_x$)$_{12}$ systems.
Following discussion in Ref. \cite{Herring51},
one can derive the approximate relation,
$\hbar\omega \approx
(A_x q_a^2 + A_x q_b^2 + A_z q_c^2)
\,4/\rho_{M}$,
where $\rho_{M}$ is the number of Bohr magnetons
per unit volume.
By comparing it with $\hbar\omega \approx
(D_a q_a^2 + D_a q_b^2 + D_c q_c^2)$,
one can see that 
\begin{equation}
 A_x = \rho_M D_a /4;\ 
 A_z = \rho_M D_c /4.
 \label{AD_prop}
\end{equation}
In the same manner we derived
$D = (D_{a}^2 D_{c})^{1/3}$,
exchange-stiffness $A$ can be related to $A_x$ and $A_z$ 
as follows:
\begin{equation}
A=(A_x^2A_z)^{1/3}.
\end{equation}
For a general $\mathcal{A}$ this relation becomes 
$A=(\lambda_1\lambda_2\lambda_3)^{1/3}$, where $\lambda$'s are 
the eigenvalues of $\mathcal{A}$.

The calculated values of $A$, $A_x$ and $A_z$ for
Sm(Fe$_{1-x}$Co$_x$)$_{12}$
are shown in Table~\ref{table2}.
The anisotropy in the spin-wave stiffness
directly leads to anisotropy in the exchange stiffness
because the relation  $D_a/D_c = A_x / A_z$ holds
due to eq. \eqref{AD_prop}.
Toga et al. \cite{Toga18} has recently reported the values of
the exchange stiffness
at finite temperatures in their Monte Carlo simulation
for Nd$_2$Fe$_{14}$B, which is also an Fe-rich rare-earth compound.
Though they did not present the value at zero temperature, 
the order of the exchange stiffness seem to agree with our results for SmFe$_{12}$.
\begin{table}[!t]
\renewcommand{\arraystretch}{1.3}
\caption{Calculated values of the exchange stiffness for
 Sm(Fe$_{1-x}$Co$_{x}$)$_{12}$.
 The stiffness
 values are in the unit of $\mathrm{pJ/m}(=10^{-12}\mathrm{J/m})$.}
 \label{table2}
 \centering
 \begin{tabular}{rcccc}
  \hline
  \hline
  $x$  & $A$ & $A_{x}$ & $A_{z}$ \\
  \hline
    0 & 7.21 & 5.61 & 11.9 \\
  0.1 & 15.6 & 13.6 & 20.5 \\
  0.2 & 20.3 & 18.1 & 25.4 \\
 \hline
\hline
 \end{tabular}
\end{table}

\section{Conclusion}
We presented calculation of the spin-wave dispersion in SmFe$_{12}$ and
Sm(Fe$_{0.8}$Co$_{0.2}$)$_{12}$, and discussed how the introduction of Co enhances the
exchange interaction. Our calculation predicted anisotropy in the
curvatures of the lowest branch around the $\Gamma$ point in SmFe$_{12}$ and
weakening of the anisotropy when Co is introduced into the system. We
also calculated the spin-wave stiffness of the systems, which values are
in good agreement with the experimental values by Hirayama et al
\cite{Hirayama17}.

\section*{Acknowledgment}

The authors would like to thank Yusuke Hirayama and Yuta Toga for sound advice
and fruitful discussions.
We gratefully acknowledge the support from the Elements Strategy Initiative Project under the auspices of MEXT. This work was also supported by MEXT as a social and scientific priority issue (Creation of new functional Devices and high-performance Materials to Support next-generation Industries; CDMSI) to be tackled by using the post-K computer. 
The computation was partly conducted using the facilities of the Supercomputer Center, the Institute for Solid State Physics, the University of Tokyo, and the supercomputer of ACCMS, Kyoto University. 
This research also used computational resources of the K computer provided by the RIKEN Advanced Institute for Computational Science through the HPCI System Research project (Project ID:hp170100). 

\appendix
\section{Heisenberg Hamiltonian for spin-waves}
\label{Appendix_1}
In this section, we consider small fluctuation
of the spins from the ground state
in the Hamiltonian of eq. \eqref{Heisenberg_1}.
We assume the ground state to be the form of 
$\vec{e}^{\,\text{GS}, \mu}_i = {}^{t}(0, 0, \sigma_i)$
with $\sigma_i = \pm 1$, which means the spins are parallel or antiparallel
along the
$z$-axis and the unit cell is large enough to describe the ground state.
We assume the $x$-component of the fluctuation, $\delta x^{\mu}_{i}$, 
is much smaller than 1, and so is the $y$-component, $\delta y^{\mu}_{i}$.
With this fluctuation, the $\vec{e}$ can be written as follows:
\begin{equation}
 \vec{e}^{\,\mu}_i
  \simeq \left(
  \begin{array}{c}
      \delta x^{\mu}_{i} \\
      \delta y^{\mu}_{i} \\
      \sigma_i
      \left[
          1-\frac{1}{2}
          \left\{
	   (\delta x^{\mu}_{i})^2+(\delta y^{\mu}_{i})^2
	  \right\}
      \right]
  \end{array}
  \right).
\end{equation}
We here retain the second order terms of $\delta x$ and $\delta y$
because the excitation energy is the second order of the fluctuations.
It can easily be seen that this vector has an absolute value of 1
within the second order approximation.
Then, the Hamiltonian of eq. \eqref{Heisenberg_1} becomes
\begin{equation}
 H - E^\mathrm{GS}
  =  - \sum_{\alpha \in \{x, y\}}
  \sum_{i,\mu} \sum_{j,\nu}\,
  \delta\alpha^{\,\mu}_i
  \left(
      J^{\mu,\nu}_{i,j} - \delta_{\mu,\nu}\sum_{k,\iota} J^{\mu,\iota}_{i,k}
  \right)
  \delta\alpha^{\,\nu}_j,
\label{Heisenberg_2}
\end{equation}
where $E^\mathrm{GS}$ is the energy of the ground state,
and $\delta\alpha$ is either $\delta x$ or $\delta y$ indexed by $\alpha$
in the first summation symbol.
This expression can be rewritten with the Fourier transform of 
$\delta \alpha$ and $J$, 
$
\delta \alpha^{\,\mu}_i =
\sum_{\vec{q}}
\delta \tilde{\alpha}^{\,\mu}(\vec{q})
\,e^{-i\vec{q}\cdot\vec{R}_i}
$
and 
$
J^{\,\mu,\nu}_{i,j} =
\sum_{\vec{q}}
\tilde{J}^{\,\mu,\nu}(\vec{q})
\,e^{-i\vec{q}\cdot\left(\vec{R}_i-\vec{R}_j \right)}
$,
into the following form:
\begin{align}
 H - E^\mathrm{GS}
  &=  - \sum_{\alpha \in \{x, y\}}
  \sum_{\mu,\nu}
  \,
  \overline{\delta\tilde{\alpha}^{\,\mu}(\vec{q})}
  \left(
      \tilde{J}^{\mu,\nu}(\vec{q})
      -
      \delta_{\mu,\nu}\sum_{\iota} \tilde J^{\mu,\iota}(\vec{0})
  \right)
  \,
  \delta\tilde{\alpha}^{\,\nu}(\vec{q})
\label{Heisenberg_3} \\
  &\equiv  - \sum_{\alpha \in \{x, y\}}
  \sum_{\mu,\nu}
  \,
  \overline{\delta\tilde{\alpha}^{\,\mu}(\vec{q})}
  \,
  \tilde{\mathcal{J}}^{\mu, \nu}(\vec{q})
  \,
  \delta\tilde{\alpha}^{\,\nu}(\vec{q})
\label{Heisenberg_4}
\end{align}
because $\delta\alpha^{\,\mu}_i$ is real.
Therefore, we can obtain elementary excitations in the Hamiltonian
by diagonalizing the matrix of
$\tilde{\mathcal{J}}^{\mu,\nu}(\vec{q})=
      \tilde{J}^{\mu,\nu}(\vec{q})
      -
      \delta_{\mu,\nu}\sum_{\iota} \tilde J^{\mu,\iota}(\vec{0})
$.

\section{Spin-wave stiffness in bcc Fe, hcp Co, fcc Co and fcc Ni}
\label{Appendix_B}
In this section, we show values of the spin-wave stiffness calculated for
bcc Fe, hcp Co, fcc Co and fcc Ni.

We used calculated values of the lattice constants
of
$a=2.837$~\AA\ for\ bcc Fe,
$a=2.494$~\AA, $c=4.042$~\AA\ for hcp Co,
$a=3.530$~\AA\ for fcc Co,
and
$a=3.525$~\AA\ for fcc Ni,
which are obtained by using a PAW-GGA method\cite{QMAS}.
The spin-wave stiffness are determined by fitting of a quadratic function
to the data points within
a $0.2 \times \frac{2\pi}{a}$ radius (bcc, hcp)
or
a $0.3 \times \frac{2\pi}{a}$ radius (fcc)
of the $\Gamma$ point.

Table \ref{table_B1} compares our values with previous theoretical values \cite{Pajda01} and experimental values.
Our results are in good agreement with the experimental values except for Ni, for which the Heisenberg model would not be very appropriate.
They are also quite similar to the previous theoretical values, however, we could not resolve the large deviation in fcc Co.

\begin{table}[!t]
 \renewcommand{\arraystretch}{1.3}
 \caption{Calculated values of the spin-wave stiffness for
 bcc Fe, hcp Co, fcc Co and fcc Ni
 compared with theoretical values from \cite{Pajda01} and
 experimental values \cite{Hasegawa79,Pauthenet82b,Shirane68}.
 The values are in the unit of $\mathrm{meV \AA}^2$.}
 \label{table_B1}
 \centering
 \begin{tabular}{cccc}
  \hline
  \hline
     & This work & Theory\cite{Pajda01} & Exp. \\
  \hline
  bcc Fe & 259 & $250 \pm 7$  & $^\mathrm{a}278\pm 30$, $^\mathrm{b}$280\\
  hcp Co & 580 & ---  & $^\mathrm{b}$580\\
  fcc Co & 429 & $663 \pm 6$  & $^\mathrm{c}$340 \\
  fcc Ni & 748 & $756 \pm 29$ & $^\mathrm{d}$555\\
 \hline
\hline
 \multicolumn{4}{l}{$^\mathrm{a}$ Ref. \cite{Hasegawa79}; $^\mathrm{b}$ Ref. \cite{Pauthenet82b}; $^\mathrm{d}$ Ref. \cite{Mook73}}   \\
 \multicolumn{4}{l}{$^\mathrm{c}$ By thin-film resonance at 295 K \cite{Shirane68}} \\
 \end{tabular}
\end{table}

\section{The $D_a/D_c$ ratio in bcc Fe, fcc Co and fcc Ni with
 tetragonal deformation}
\label{Appendix_C}
To provide more data of the anisotropy of the spin-wave stiffness
in tetragonal systems
we calculate the $D_a/D_c$ ratio in bcc Fe, fcc Co and fcc Ni with
applying tetragonal deformation.
The $c/a$ ratio of the systems is changed from -2\% to +2\%
with their volume kept constant
(see \ref{Appendix_B} for the systems without the deformation).

The $D_a/D_c$ ratio can be approximated by $(c/a)^{-2}$
when the change of $J^{\mu, \nu}_{i,j}$ 
caused by the deformation can be disregarded.
Consider a linear regular transform $T$ for the lattice vectors, 
$\vec{R} \rightarrow \vec{R}' = T \vec{R}$.
The tetragonal deformation we are considering here
can be expressed by 
$T =
\text{diag}\left[
\sqrt[3]{a/c}, \sqrt[3]{a/c}, \sqrt[3]{(c/a)^2}
\right]$.
This transforms a vector $\vec{q}$
in the reciprocal space to $\vec{q}\,' = {}^t(T^{-1})\,\vec{q}$,
which keeps $\vec{q}\cdot \vec{R}$ invariant.
With the fixed values of $J^{\mu, \nu}_{i,j}$
we can obtain 
$
 \tilde{J}'^{\,\mu,\nu}({\vec q}\,')
  =
 \tilde{J}^{\,\mu,\nu}({\vec q})
$
by the inverse Fourier transform.
This leads to the approximation
\begin{equation}
 \tilde{\mathcal{J}}'^{\,\mu,\nu}({\vec q}\,')
  \simeq
 \tilde{\mathcal{J}}^{\,\mu,\nu}({\vec q})
\end{equation}
Note that the Fourier transform with respect to the lattice after the operation of $T$ (for which we put a prime) must be distinguished from that with respect to the original lattice.

Therefore, the approximate relation in the cubic system
$\hbar\omega \approx D ({\vec q}\,_a^2+{\vec q}\,_b^2+{\vec q}\,_c^2)$
is transformed into the form of 
$\hbar\omega \approx D'_a {\vec q}\,_a'^2+ D'_b {\vec q}\,_b'^2+D'_c {\vec q}\,_c'^2$
with $D'_a = D'_b = (c/a)^{-2/3}D$ and $D'_c = (c/a)^{4/3}D$.
Finally, we obtain the approximate relation $D'_a/D'_c = (c/a)^{-2}$.

\begin{figure}[!t]
\centering
 \includegraphics[width=3.5in,bb=0 0 504 360]{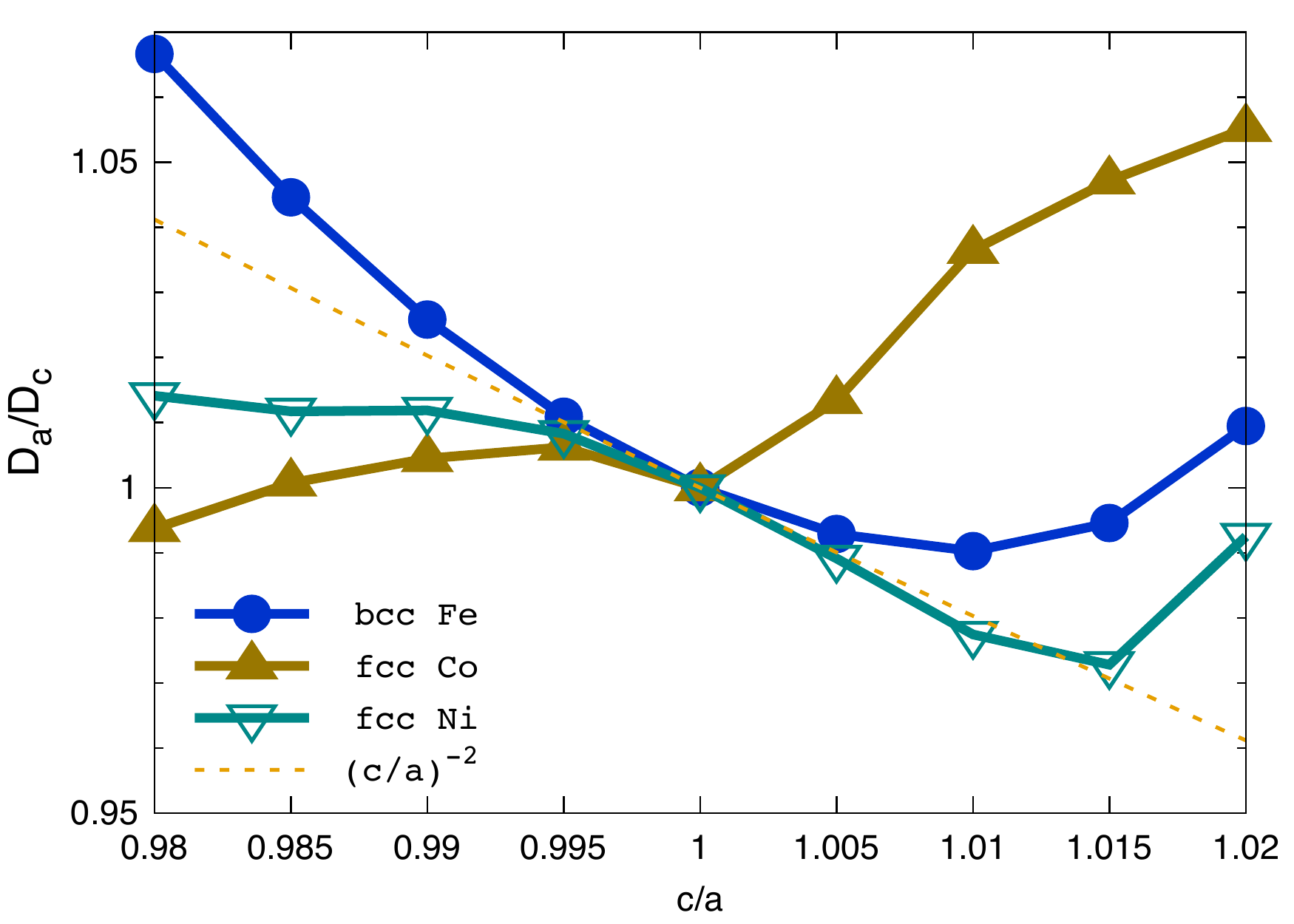}
\caption{
The $D_a/D_c$ ratio for bcc Fe, fcc Co, and fcc Ni with
tetragonal distortion as functions of the $c/a$ ratio. 
The volumes are fixed to those of the cubic systems ($c/a=1$)
in the calculations.
The dotted line shows the function of $(c/a)^{-2}$.
}
\label{fig_B2}
\end{figure}

Figure \ref{fig_B2} shows the $D_a/D_c$ ratio for bcc Fe, fcc Co, and fcc Ni
with the tetragonal deformation.
The $D_a/D_c$ ratio in Fe and Ni tend to have a negative correlation with $c/a$,
which is opposite to the cases of SmFe$_{12}$ (this work)
and Nd$_2$Fe$_{14}$B \cite{Toga18}.
In contrast, fcc Co has a positive correlation to $c/a$.
These can be attributed to a difference in dependency of
$J^{\mu, \nu}_{i,j}$ on $c/a$, which leads to the deviation from 
the $(c/a)^{-2}$ function.

\bibliography{spinwave}
\bibliographystyle{unsrt}

\end{document}